\documentclass[%
 ,twocolumn%
 ,secnumarabic%
,amssymb, amsmath,nobibnotes, aps, prl]{revtex4}
\usepackage{epsfig}%
\usepackage{docs}%
\usepackage{bm}%
\expandafter\ifx\csname package@font\endcsname\relax\else
 \expandafter\expandafter
 \expandafter\usepackage
 \expandafter\expandafter
 \expandafter{\csname package@font\endcsname}%
\fi

\begin{document}

\def\G{GeV$^2$}
\def\Q{$Q^2$}
\def\v{\vspace{.1in}}
\def\F{\mathcal{F}}
\def\K{\mathcal{K}}

\title{\revtex~4 Connection between the elastic $G_{Ep}/G_{Mp}$ and $\rm P\to \Delta$ form factors.}%

\author{Paul Stoler}%
\email{stoler@rpi.edu}
\affiliation {Physics Department, Rensselaer Polytechnic Institute, Troy, NY 12180}
\date{October 11, 2002}%

\begin{abstract}
It is suggested that the falloff in \Q\ of the  $\rm P\to \Delta$ magnetic form
factor $G^*_M$ is related to the recently observed falloff of the
elastic electric form factor $G_{Ep}/G_{Mp}$. Calculation is
carried out in the framework of a GPD model whose
parameters are determined by fitting the elastic form factors
$F_{1p}$ and $F_{2p}$ and isospin invariance. When applied to the 
$P\to \Delta$ transition with no additional parameters, the shape of $G^*_M$  
is found to exhibit the requisite falloff with \Q . 

\end{abstract}

\maketitle

The 
$P\to \Delta(1232)$ form factor $G^*_M$ 
exhibits a more rapid decrease with respect
to \Q\ than is typically observed in
 other baryons~\cite{stoler}~\cite{frolov}, such as $G_{Mp}$ in
elastic scattering from a proton, or $A_{1/2}$
in the transition $P\to S_{11}(1535)$. 
A recent  Jefferson Lab (JLab) measurement~\cite{perdrisat}
finds that the ratio $G_{Ep}/G_{Mp}$ for elastic scattering
falls with \Q\ more rapidly than previously
expected. This has given rise to much theoretical 
activity~\cite{miller}~\cite{ralston}
to attempt to understand the underlying physics.
In this note it is suggested that this behavior in 
$G_{Ep}/G_{Mp}$ is related to that of $G^*_M$. 

As a basis it is assumed that the form factor is dominated by
soft mechanisms, and a 
GPD-handbag approach~\cite{ji}~\cite{rad_gpd}~\cite{collins} is
utilized. Form factors  are  the zero'th  moments of the GPDs 
with skewedness $\xi$ = 0. For elastic 
scattering, the Dirac and Pauli form factors are given by 
 
\begin{equation}
F_1(t)=\int^1_{-1}\sum_q  H^q(x,\xi, t)dx  
\label{eq:F1}
\end{equation}
\begin{equation}
F_2(t)=\int^1_{-1}\sum_q  E^q(x,\xi, t)dx
\label{eq:F2}
\end{equation}

\noindent where $q$ signifies flavors.
In the following, with $\xi $=0, for brevity the GPD's are denoted
$ H^q(x,t) \equiv H^q(x,0,t)$, and \ $ E^q(x,t) \equiv E^q(x,0,t)$.

Resonance transition form factors access components of the
GPDs which are not accessed in elastic scattering.
The  $N\to\Delta$ form factors are related
to isovector components of the GPDs~\cite{frankfurt}~\cite{goeke}; 

%\begin{widetext}
\begin{eqnarray}
2G^*_M & = & \int^1_{-1}\sum_q H^q_M(x,t)dx \\
2G^*_E & = &\int^1_{-1}\sum_q H^q_E(x,t)dx \\  
2G^*_C & = &\int^1_{-1}\sum_q H^q_C(x,t)dx 
\label{eq:delta}
\end{eqnarray}
%\end{widetext}

\noindent where  $G^*_M$, $G^*_E$ and  $G^*_C$ are magnetic, electric
and Coulomb transition form factors~\cite{jones}, and 
$ H^q_M$,  $H^q_E$, and $H^q_C$ are isovector GPDs,
which can be related to elastic GPDs in the large $N_C$ chiral limit
through isospin rotations. Analogous relationships can be obtained 
for the $N\to S_{11}$ and other  transitions.
Here, the connection between GPDs involved in
the elastic and $N\to\Delta$ form factors is explored to
obtain the connection between the \Q\ dependence of the
$G_{Ep}$ and $G^*_M$.

In refs.~\cite{frankfurt}~\cite{goeke} it is noted that, in the large $N_c$
limit, assuming chiral and isospin symmetry
the GPDs for the $P\to \Delta(1232)$ transition are expected  to be
isovector components of the  elastic GPDs,  given by

\begin{equation}
H^{(IV)}_M = {2\over{\sqrt{3}}}E^{(IV)}_M={2\over{\sqrt{3}}}
\left(E_p^u-E_p^d\right).
\label{eq:IV}
\end{equation}

\noindent $E_p^u$ and $E_p^d$ are the GPD's for the $u$ and $d$
quarks respectively. Thus the $P\to \Delta$ form factor should be
obtainable by analysis of the Pauli form factor
 $F_{2p}$ (eq.~\ref{eq:F2}).  The Dirac and Pauli form factors,
$F_{1p}$ and $F_{2p}$,  
are related to the measured Sachs form factors $G_{Mp}$ and $G_{Ep}$ 
by

\begin{eqnarray}
F_{1P}(Q^2)={1\over{\tau +1}}(\tau G_{Mp}(Q^2)+ G_{Ep}(Q^2))\\
F_{2P}(Q^2)={1\over{\kappa(\tau +1)}}(G_{Mp}(Q^2)- G_{Ep}(Q^2))\\
\label{eq:GF}
\end{eqnarray}

\noindent with $\tau=Q^2/4M_p$. To obtain  $E_p^u$ and
$E_p^d$,  needed for eq.~\ref{eq:IV}, the available data for $G_{Mp}$ and 
the recent JLab data~\cite{perdrisat}  on $G_{Ep}/G_{Mp}$ were fit,
as reported in ref.~\cite{stoler_gpd}, using a parameterization of the 
GPDs such as in~\cite{rad_wacs}~\cite{kroll}~\cite{afanasev}~\cite{burkardt}.

The specific functional form for $H_P(x,t)$  and $E_P(x,t)$  is
a Gaussian plus small power law shape in $-t\ (\equiv Q^2)$ to account for
the high the measured form factors at very high \Q.

\begin{eqnarray} 
H_P(x,t)=f(x)exp(\bar x t /4x\lambda_H^2) + \cdot\cdot\cdot \\
E_P(x,t)=k(x)exp(\bar x t /4x\lambda_E^2) +  \cdot\cdot\cdot,
\end{eqnarray}

\noindent in which $\bar x \equiv 1-x$, and  $\cdot\cdot\cdot$ indicates
the addition of small power components added in ref.~\cite{stoler_gpd}
to account for higher $Q^2$ contributions to the form factors.
The  conditions at \Q=0 are $H(x,0)=e_uf_u(x)+e_df_d(x)$ and 
$E(x,0)=k_u(x)+k_d(x)$. Here, $f_u(x)$ and $f_d(x)$ are the $u$ and $d$
valence quark distribution functions measured in DIS.
The functions $k_{u}(x)$ and $k_{d}(x)$ are not obtainable 
from DIS. Following ref.~\cite{afanasev} the form used was
$k^q(x)\propto\sqrt{1-x}f^q(x)$, with normalizatons obtained
employing isospin symmetry, and  by requiring the
proton and neutron form factors to have their known values
near \Q =0, that is  $F_{1p}(0)=1$, $F_{2p}(0)=1.79$, $F_{1n}(0)=0$,
$F_{2n}(0)=-1.91$. This gives 

\begin{eqnarray}
F_{2u}(0)\equiv \kappa_u = \int{k_u(x)dx} = 1.67 \\
F_{2d}(0)\equiv \kappa_d = \int{k_d(x)dx} = -2.03 
\label{eq:F2ud}
\end{eqnarray}

\noindent and

\begin{eqnarray}
F_{1u}(0)= \int{e_uf_u(x)dx} = 2/3 \\
F_{1d}(0)= \int{e_df_d(x)dx} = 1/3
\end{eqnarray}

\begin{figure}[htb]
\centerline{\psfig{file=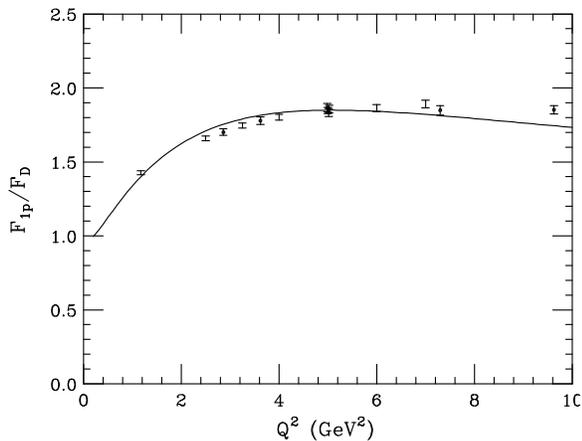,angle=90,width=3in}}
\caption{\label{fig4} Dirac form factor
$F_{1p}(Q^2)$ relative to the dipole  $G_D=1/(1+Q^2/.71)^2$.
The data are extracted  using the recent JLab. data~\cite{perdrisat}
for $G_{Ep}/G_{Mp}$, and a recent reevaluation~\cite{brash}
of SLAC data of $G_{Mp}$~\cite{arnold}~\cite{andivahis}. 
The curve is the result  of the fit as discussed in the text.}
\end{figure}

\begin{figure}[htb]
\centerline{\psfig{file=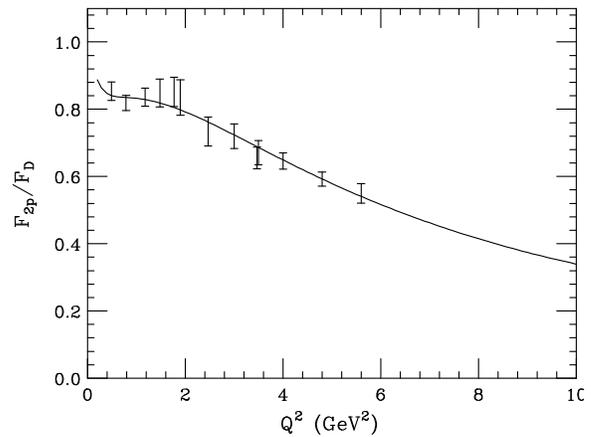,angle=90,width=3in}}
\caption{
\label{fig5} The Pauli form factor
$F_{2}/1.79F_D$ relative to the dipole  $F_D=1/(1+Q^2/.71)^2$.
The data are extracted using the recent JLab. data~\cite{perdrisat}
for $F_{2p}/F_{1p}$, multiplied by the fit curve for $F_{1p}/F_D$
shown in fig.~\ref{fig4}. The curve is the result 
of the simultaneous fit to the  $G_{Ep}/G_{Mp}$ and $G_{Mp}$
data as discussed in the text and fig.~\ref{fig4}  . }
\end{figure}

\noindent Adequate fits to $G_{Mp}$ and $G_{Ep}/G_{Mp}$, or
equivalently $F_{1p}$ and  $F_{2p}/F_{1p}$, were otained
with $\lambda_H=0.76$ GeV/c and $\lambda_E=0.67$ GeV/c.
The results are shown in figs.~\ref{fig4} and \ref{fig5}.

\begin{figure}[h]
\centerline{\psfig{file=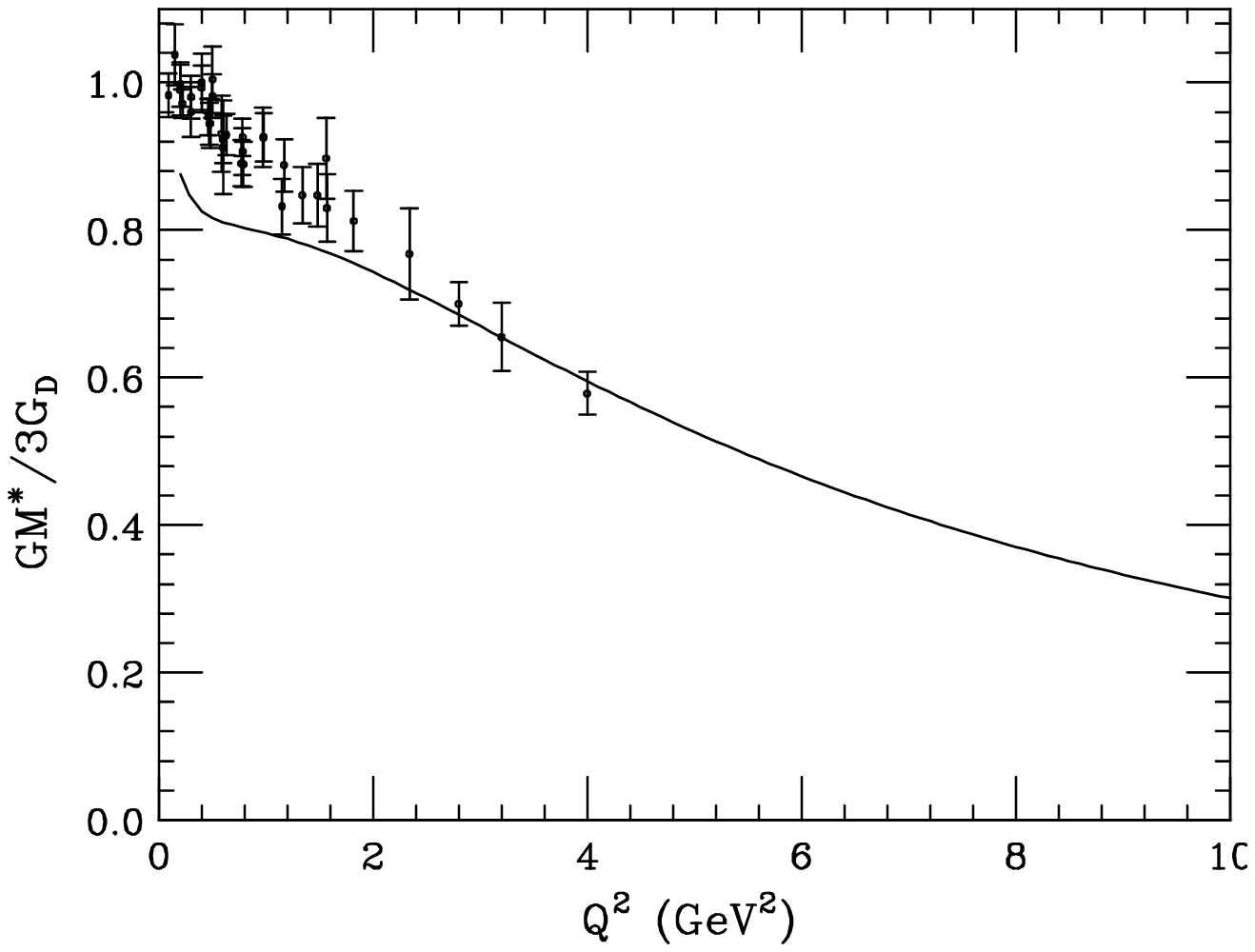,angle=90,width=3in}}
\caption{ \label{fig6} The $N\to\Delta$ magnetic form factor
$G_M^*(Q^2)/3G_D$ relative to the dipole  $G_D=1/(1+Q^2/.71)^2$ }.
The data are a compendium of world data by ref~\cite{kamalov}.
The curve is the result of the procedures discussed in the text.
\end{figure}

The resulting $E^u_p$ and $E^d_p$ were inserted into
eq.~\ref{eq:IV} to obtain an estimate for $G^*_M$. At
\Q=0 one gets $G^*_M(0)=2.14$, which is somewhat lower than
the experimental value of  $G^*_M(0)\sim 3$. Such a disagreement 
is not surprising~\cite{frankfurt}~\cite{goeke}  given the very 
approximate nature of  eq.~\ref{eq:IV}. The obtained $G^*_M$ 
was  overall renormalized to take this ratio
into account, and the result is shown in fig.~\ref{fig6}.

The similar shapes of the curves in figs.~\ref{fig5} and ~\ref{fig6}
can be ascribed to their connection via eq.~\ref{eq:IV}. This can
be understood by the observation that $F_2$ is nearly all isovector
spin-flip, as is the $G_M^*$. However, the difference in the mass of
the $\Delta(1232)$ and the nucleon, which  is a measure of the SU3 
symmetry breaking, and the fact that $F_1$ also has an isovector
component would make the observed non-negligible differences in the
normalization not surprising. 

Although this note suggests a common physical origin in the
the \Q\ behavior of  $G_{Ep}/G_{Mp}$ and $G_M^*$, a complete
understanding  will require theoretical treatments
based on rigorous and consistent relativistic treatment which are beyond
the scope of this communication.

{\bf Acknowlegments:} The author thanks G.A. Miller, A.V. Radyushkin and
M. Vanderhaeghen for helpful discussions.  
The work was partially supported by the {\em National Science Foundation}.

\end{document}